\documentclass[twocolumn]{aastex701}
\usepackage{color, hyperref}
\usepackage{amsmath}
\usepackage{xspace}
\usepackage{lmodern}

\newcommand{\hf}{HF(1--0)\xspace}
\newcommand{\f}{$^{19}$F\xspace}

\begin{document}

\title{Upper limit on \hf absorption in a dusty star-forming galaxy at $z = 6$:\\ Constraints on early fluorine enrichment}

\author[0000-0002-0498-5041]{Akiyoshi Tsujita}
\affiliation{Institute of Astronomy, Graduate School of Science, The University of Tokyo, 2-21-1 Osawa, Mitaka, Tokyo 181-0015, Japan}
\email[show]{tsujita@ioa.s.u-tokyo.ac.jp}  

\author[0000-0002-4343-0487]{Chiaki Kobayashi}
\affiliation{Kavli Institute for the Physics and Mathematics of the Universe (WPI), The University of Tokyo Institutes for Advanced Study, The University of Tokyo, Kashiwa, Chiba 277-8583, Japan}
\affiliation{Centre for Astrophysics Research, Department of Physics, Astronomy and Mathematics, University of Hertfordshire, Hatfield, AL10 9AB, UK}
\email{c.kobayashi@herts.ac.uk}

\author[0000-0002-1413-1963]{Yuki Yoshimura}
\affiliation{Institute of Astronomy, Graduate School of Science, The University of Tokyo, 2-21-1 Osawa, Mitaka, Tokyo 181-0015, Japan}
\email{}

\author[0000-0002-4052-2394]{Kotaro Kohno}
\affiliation{Institute of Astronomy, Graduate School of Science, The University of Tokyo, 2-21-1 Osawa, Mitaka, Tokyo 181-0015, Japan}
\affiliation{Research Center for the Early Universe, Graduate School of Science, The University of Tokyo, 7-3-1 Hongo, Bunkyo-ku, Tokyo 113-0033, Japan}
\email{kkohno@ioa.s.u-tokyo.ac.jp}

\author[0000-0001-9728-8909]{Ken-ichi Tadaki}
\affiliation{Faculty of Engineering, Hokkai-Gakuen University, Toyohira-ku, Sapporo 062-8605, Japan}
\email{tadaki@hgu.jp}

\author[0000-0002-8868-1255]{Fumiya Maeda}
\affiliation{Research Center for Physics and Mathematics, Osaka Electro-Communication University, 18-8 Hatsucho, Neyagawa, Osaka 572-8530, Japan}
\email{fmaeda@osakac.ac.jp}

\author[0000-0003-1937-0573]{Hideki Umehata}
\affiliation{Institute for Advanced Research, Nagoya University, Furocho, Chikusa, Nagoya 464-8602, Japan}
\affiliation{Department of Physics, Graduate School of Science, Nagoya University, Furocho, Chikusa, Nagoya 464-8602, Japan}
\email{umehata@a.phys.nagoya-u.ac.jp}

\author[0009-0006-1731-6927]{Shuo Huang}
\affiliation{Department of Physics, Graduate School of Science, Nagoya University, Furocho, Chikusa, Nagoya 464-8602, Japan}
\affiliation{National Astronomical Observatory of Japan, 2-21-1 Osawa, Mitaka, Tokyo 181-8588, Japan}
\email{shuohuang.version3@gmail.com}

\author[0000-0001-6469-8725]{Bunyo Hatsukade}
\affiliation{National Astronomical Observatory of Japan, 2-21-1 Osawa, Mitaka, Tokyo 181-8588, Japan}
\affiliation{Department of Astronomy, Graduate School of Science, The University of Tokyo, 2-21-1 Osawa, Mitaka, Tokyo 181-0015, Japan}
\affiliation{Graduate Institute for Advanced Studies, SOKENDAI, Osawa, Mitaka, Tokyo 181-8588, Japan}
\email{bunyo.hatsukade@nao.ac.jp}

\author[0000-0002-1639-1515]{Fumi Egusa}
\affiliation{Institute of Astronomy, Graduate School of Science, The University of Tokyo, 2-21-1 Osawa, Mitaka, Tokyo 181-0015, Japan}
\email{fegusa@ioa.s.u-tokyo.ac.jp}

\author[0000-0003-3932-0952]{Kana Morokuma-Matsui}
\affiliation{Institute of Astronomy, Graduate School of Science, The University of Tokyo, 2-21-1 Osawa, Mitaka, Tokyo 181-0015, Japan}
\email{kanamoro@ioa.s.u-tokyo.ac.jp}

\author[0000-0003-4807-8117]{Yoichi Tamura}
\affiliation{Department of Physics, Graduate School of Science, Nagoya University, Furocho, Chikusa, Nagoya 464-8602, Japan}
\email{ytamura@nagoya-u.jp}

\author[0000-0003-0563-067X]{Yuri Nishimura}
\affiliation{Institute of Pure and Applied Sciences, University of Tsukuba, 1-1-1 Tennodai, Tsukuba, Ibaraki, 305-8577, Japan}
\affiliation{Tomonaga Center for the History of the Universe, University of Tsukuba, 1-1-1 Tennodai, Tsukuba, Ibaraki, 305-8577, Japan}
\affiliation{Tsukuba Institute for Advanced Research (TIAR), University of Tsukuba, 1-1-1 Tennodai, Tsukuba, Ibaraki, 305-8577, Japan}
\email{nishimura.yuri.ke@u.tsukuba.ac.jp}

\begin{abstract}
Wolf-Rayet (WR) stars have recently attracted attention as possible drivers of early chemical enrichment, including the production of fluorine, whose nucleosynthetic origin remains debated.
To test the contribution of massive stars to fluorine production in the early Universe, we conducted Atacama Large Millimeter/submillimeter Array Band~5 spectroscopy of the \hf absorption line toward a dusty star-forming galaxy at $z=6.024$. This galaxy has a known gas-phase metallicity and is too young for low-mass AGB stars to have contributed significantly, providing a clean environment to isolate massive-star yields. 
We do not detect significant HF absorption ($\sim2\sigma$) and derive a conservative 5$\sigma$ upper limit of $N_\mathrm{HF}/N_\mathrm{H_2} < 2.2\times10^{-9}$.
This limit is about an order of magnitude below typical local measurements, indicating inefficient fluorine enrichment $\sim0.9$\,Gyr after the Big Bang.
Comparison with chemical evolution models shows that our constraint is consistent with scenarios without WR yields at this epoch. 
Expanding the sample of HF absorption measurements in high-redshift galaxies with well-characterized metallicities will be crucial for tracing the onset of WR enrichment and fluorine production across cosmic time.
\end{abstract}



\section{Introduction} 
Recent discoveries of unusually high N/O ratios in high-redshift galaxies have drawn attention to Wolf–Rayet (WR) stars as a possible driver of early chemical enrichment \citep[e.g.,][]{Bunker2023A&A...677A..88B, Isobe2023ApJ...959..100I, Kobayashi2024ApJ...962L...6K, Berg2025arXiv251113591B}.
Another element for which WR stars have been suggested to contribute to its production is fluorine \citep{Franco2021NatAs...5.1240F}.
Fluorine (\f) is one of the few elements whose nucleosynthetic origin and evolution are still much debated. 
Its unusually low cosmic abundance compared to neighboring elements indicates that standard stellar nucleosynthesis largely bypasses fluorine, making it a sensitive probe of non-standard stellar yields \citep[e.g.,][]{Ryde2020ApJ...893...37R}.
Theoretically, three astrophysical objects have been considered as the primary sources of \f production: asymptotic giant branch (AGB) stars, WR stars, and core-collapse supernovae (SNe II) with the neutrino process.
Among them, AGB stars are the only observationally confirmed producers \citep[e.g.,][]{Jorissen1992A&A...261..164J, Werner2005A&A...433..641W, Otsuka2008ApJ...682L.105O, Lucatello2011ApJ...729...40L, Abia2010ApJ...715L..94A, Abia2015A&A...581A..88A}, and low-mass AGB stars ($\sim2$--4 $M_\odot$) are considered the main contributors to \f production at solar metallicity \citep[e.g.,][]{Lugaro2004ApJ...615..934L, Karakas2010MNRAS.403.1413K, Kobayashi2011MNRAS.414.3231K, Womack2023MNRAS.518.1543W}.
However, AGB yields alone cannot reproduce the observed fluorine abundances in the solar neighborhood, implying additional contributions from massive stars such as WR stars and SNe II \citep[e.g.,][]{Renda2004MNRAS.354..575R, Kobayashi2011ApJ...739L..57K, Kobayashi2020ApJ...900..179K, Jonsson2014A&A...564A.122J, Spitoni2018A&A...612A..16S, Prantzos2018MNRAS.476.3432P, Ryde2020ApJ...893...37R}.
Although non-AGB contributions are expected, direct tests are scarce mainly because fluorine tracers are limited to the vibrational–rotational HF lines in the infrared and the rotational transitions in the millimeter regime, the latter being inaccessible to local stars.
The vibrational-rotational HF lines are difficult to detect in extremely metal-poor stars, limiting stellar fluorine measurements to a small number of carbon-enhanced objects \citep[e.g.,][]{Mura-Guzman2020MNRAS.498.3549M}. 
This is primarily because most metal-poor stars have sufficiently high effective temperatures to dissociate HF molecules.
Moreover, such observations probe the photospheric \f abundance of individual stars rather than the element already ejected into the interstellar medium (ISM), leaving the efficiency of \f ejection uncertain. 

A complementary approach is to probe fluorine in the ISM of young but already metal-rich galaxies, where AGB enrichment is still limited but short-lived massive stars can operate efficiently.
Recently, \citet{Franco2021NatAs...5.1240F} serendipitously detected the \hf absorption line in a lensed dusty star-forming galaxy (DSFG) at $z=4.4$, about 1.4~Gyr after the Big Bang, when the contribution from low-mass AGB stars is expected to be small.
Their chemical evolution models showed that this could be explained either by including yields from WR stars or by assuming a short starburst timescale ($\tau_{\rm SF} \sim 0.1$\,Gyr) even without WR contributions.
However, the lack of constraints on the galaxy’s star-formation history and metallicity left this degeneracy unresolved.
To take the next step, ideal targets for HF absorption are bright DSFGs at high redshift with metallicity measurements.
While {\it James Webb Space Telescope (JWST)} has significantly improved constraints on the mass–metallicity relation (MZR) up to $z\sim10$, especially for low- to intermediate-mass galaxies \citep[e.g.,][]{Nakajima2023ApJS..269...33N, Sanders2024ApJ...962...24S, Sarker2025ApJ...978..136S}, the relation remains poorly constrained at the high-mass end ($M_\star \gtrsim 10^{10}\,M_\odot$), although recent work has begun to explore this regime \citep[e.g.,][]{Faisst2025arXiv251016106F}.
In addition, estimating stellar masses of DSFGs is difficult due to severe dust attenuation, requiring deep {\it JWST} observations. Owing to this combination, high-redshift massive DSFGs with well-determined metallicities remain rare.

Among the few most-distant, bright DSFGs known at $z>6$, G09.83808 at $z=6.024$, corresponding to a cosmic age of only $\sim0.9$ Gyr after the Big Bang, \citep{zavala2018, Tsujita2022} provides the best opportunity to extend such studies.
This DSFG has gas-phase metallicity already enriched to $Z \approx 0.5$–$0.7\,Z_\odot$, as inferred from the [N\,\textsc{ii}]$_{205\mathrm{\mu m}}$/[O\,\textsc{iii}]$_{88\mathrm{\mu m}}$ luminosity ratio combined with far-infrared continuum flux density ratio \citep{Tadaki2022}. 
In addition, this DSFG benefits from strong lensing ($\mu\sim8-9$; \citealt{Tadaki2022}) and lies on the massive end of the star-forming main sequence \citep[][]{Zavala2022}, offering a more representative view of typical $z\sim6$ massive galaxies compared to the other two known extreme DSFGs at $z>6$, which are intrinsically much brighter ($S_{870{\rm \mu m},\mathrm{intr}}>10$ mJy) and reside in proto-cluster environments \citep{Riechers2013, Marrone2018}.
In this Letter, we present ALMA \hf absorption line observations toward G09.83808, aiming to investigate early fluorine production.

This paper is organized as follows. Section \ref{sec:observation} describes the observations and data reduction. 
Section \ref{sec:results} presents the HF absorption analysis and the
derived fluorine abundance constraint.
In Section \ref{sec:discussion}, we compare our observational constraints with a chemical evolution model. 
Throughout the paper, we adopt a cosmology with $\Omega_{\mathrm{m}}=0.3$, $\Omega_\Lambda=0.7$, and $H_0=70\,\mathrm{km\,s^{-1}\,Mpc^{-1}}$ and the Kroupa initial mass function \citep{Kroupa2001MNRAS.322..231K}. 

\section{Observations and data reduction} \label{sec:observation}
\begin{figure*}[htb]
\centering
\includegraphics[width=\textwidth]{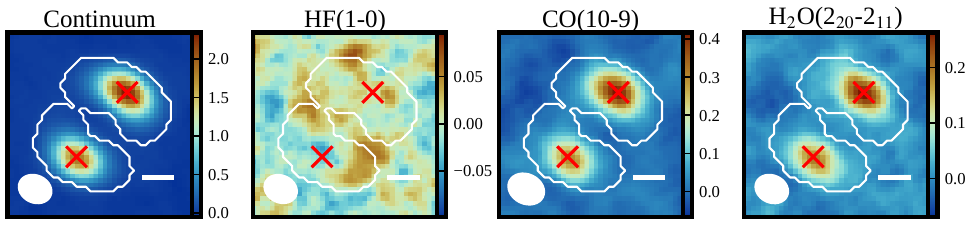}
\caption{ALMA continuum and velocity-integrated (moment-0) maps of G09.83808. From left to right: dust continuum at $\sim170$ GHz, \hf absorption, CO(10--9), and H$_2$O($2_{20}$--$2_{11}$) lines. White contours indicate the clean mask. Red crosses mark the two continuum peaks used for line spectrum extraction and photometry. The synthesized beam is shown as a white ellipse in the lower left of each panel. White scale bars indicate 1\arcsec, which corresponds to 5.7 kpc. The colorbar units are mJy\,beam$^{-1}$ for the continuum and mJy beam$^{-1}$\,km\,s$^{-1}$ for the lines.
\label{fig:mom0}}
\end{figure*}

Our ALMA Band~5 observations of G09.83808 were conducted between October and December 2024 during Cycle~10--11 (Project code: 2023.1.01281.S; PI: A. Tsujita).
We targeted the \hf absorption line ($\nu_{\rm rest}=1232.48$\,GHz; $\nu_{\rm obs}=175.47$\,GHz at $z=6.024$), covering $\sim$161.7--175.8\,GHz in the 4-bit mode.
The observations were carried out in multiple scheduling blocks using the C43-3/4/5 antenna configurations.
All datasets were calibrated using the standard ALMA pipeline in {\tt CASA} package \citep{CASA2022}. 
Continuum subtraction was performed with the \texttt{uvcontsub} task, carefully selecting line-free channels for each tuning.
We first created a clean mask from the continuum map imaged with Briggs weighting (robust parameter of 0.0), using the \texttt{auto-multithresh} algorithm \citep{Kepley2020}. This clean mask was then reused for all line cubes (white contours in Figure~\ref{fig:mom0}). 
Line imaging was carried out using the \texttt{tclean} task with natural weighting, adopting a cell size of $0.15\arcsec$ and cleaning down to \texttt{nsigma=2}. 
The final spectral cube has a synthesized beam of $1.2\arcsec \times 0.96\arcsec$ and an rms noise of 0.08\,mJy\,beam$^{-1}$ per 50\,km\,s$^{-1}$ channel at the frequency of the \hf line. Velocity-integrated maps were created by integrating over $\pm300$\,km\,s$^{-1}$ around the expected line center. 

\begin{figure}[htb]
\centering
\includegraphics[width=\columnwidth]{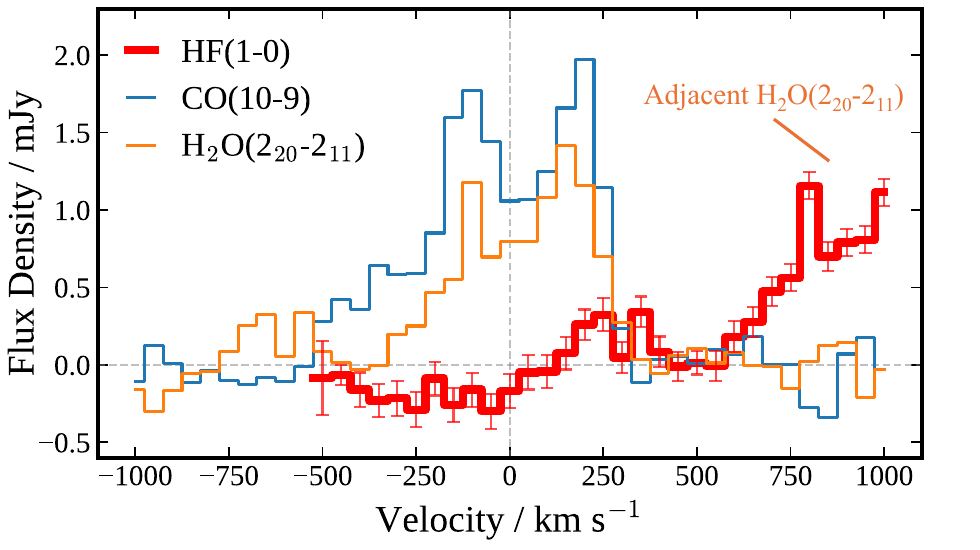}
\caption{Continuum-subtracted spectra centered on the systemic velocity. The \hf absorption line (red) is shown with $1\sigma$ error bars and binned to 50 km\,s$^{-1}$. For comparison, the CO(10--9) (blue) and H$_2$O($2_{20}$--$2_{11}$) (orange) lines from the same dataset are overlaid. 
We note that the positive feature near $+800$\,km\,s$^{-1}$ in the HF spectrum is the adjacent H$_2$O($2_{20}$--$2_{11}$) line.
\label{fig:spectrum}}
\end{figure}

Figure~\ref{fig:mom0} shows the dust continuum map and the continuum-subtracted velocity-integrated maps of the \hf, CO(10--9), and H$_2$O($2_{20}$--$2_{11}$) lines.
Given the modest angular resolution, the lensed image is resolved into two arcs in the continuum map.
We extract a single-beam spectrum at the peak pixel of each arc in the continuum image (red cross marks in Figure~\ref{fig:mom0}) and sum them channel by channel to form the total spectrum.
For each channel, the $1\sigma$ uncertainty is estimated from the rms measured in blank regions outside the source.
Figure~\ref{fig:spectrum} presents the continuum-subtracted spectrum centered on the systemic velocity, binned to 50\,km\,s$^{-1}$. For comparison, we overlay the \hf absorption line with the CO(10--9) (blended with H$_2$O($3_{12}$--$2_{21}$)) and the H$_2$O($2_{20}$--$2_{11}$) lines from the same dataset. 

\section{Results} \label{sec:results}
We calculate the fluorine abundance from the obtained spectrum.
As shown in Figure~\ref{fig:spectrum}, a tentative \hf absorption feature is seen near the systemic velocity, with a peak depth of $-0.30 \pm 0.08$\,mJy ($3.8\sigma$)  at $-50$\,km\,s$^{-1}$. 
We define the integration range using the unblended H$_2$O($2_{11}$--$2_{02}$) line, selecting channels with $>1\sigma$ signal, which yields $[-300, +300]$ km\,s$^{-1}$.
Within this range, the \hf velocity-integrated flux is $-43 \pm 20$ mJy\,km\,s$^{-1}$ ($2.2\sigma$).
Given the modest signal-to-noise ratio, we conservatively treat this measurement as an upper limit in the analysis below.
For reference, we also calculate the integrated flux over $[-450, 0]$ km\,s$^{-1}$ where the HF signal itself exceeds $1\sigma$, which yields $-97 \pm 17$ mJy\,km\,s$^{-1}$ ($5.7\sigma$).
However, using either this value or the $5\sigma$ upper limit adopted below does not affect our main conclusions.
The continuum level, measured at the same two pixels and summed in the same way, is $4.24 \pm 0.02$\,mJy. 

Because the HF $J=1$ level has an extremely high critical density of 
$n_\mathrm{crit} \sim 10^{10}\,\mathrm{cm^{-3}}$ at $T=50$\,K \citep{Neufeld2005ApJ...628..260N}, essentially all HF resides in the ground state and the $J=1$--0 absorption traces the total HF column density. 
The HF column density can therefore be obtained from the velocity–integrated optical depth 
using the relation given by \cite{Neufeld2010A&A...518L.108N}:
\begin{equation*}
\int \tau(v)\, dv = 4.16 \times 10^{-13}  N_\mathrm{HF} \quad \mathrm{cm^2\,km\,s^{-1}}
\end{equation*}
Applying this relation to the $5\sigma$ flux limit yields $N_\mathrm{HF} < 6.0 \times 10^{14}~\mathrm{cm^{-2}}\quad (5\sigma)$. 
However, we note that this upper limit may still underestimate the true HF abundance, because (1) the continuum emission may not fully illuminate all absorbing
clouds along the line of sight, (2) only gas in front of the continuum source contributes to the absorption, and (3) any HF in excited states is not accounted for.

Next, we constrain the HF abundance ratio by estimating the molecular hydrogen column density as follows.
Using the intrinsic molecular gas mass $(2.0\pm0.7)\times10^{10}\,M_\odot$ from CO(2--1) line \citep{Zavala2022}, together with a CO-to-H$_2$ conversion factor of $\alpha_{\rm CO}=1.0\,M_\odot\,(\mathrm{K\,km\,s^{-1}\,pc^2})^{-1}$, CO excitation ratio of $r_{21}=0.83\pm0.10$ \citep{Bothwell2013}, and assuming the gas follows the [C\,\textsc{ii}]$_{158\mathrm{\mu m}}$ distribution with a half-light radius of $\sim$1\,kpc (A. Tsujita et~al.~in preparation), we obtain $N_\mathrm{H_2} = 2.7 \times 10^{23}~\mathrm{cm^{-2}}$. 
The corresponding abundance limit is $N_\mathrm{HF}/N_\mathrm{H_2} < 2.2 \times 10^{-9}$ or $\log_{10}\left(N_\mathrm{HF}/N_\mathrm{H_2}\right) < -8.7 \quad (5\sigma)$.
We adopt a $\alpha_{\rm CO}$ value commonly used for local ULIRGs \citep[e.g.,][]{Downes1998ApJ...507..615D}. 
For G09.83808, dynamical constraints suggest $\alpha_{\rm CO}<2.5\,M_\odot\,(\mathrm{K\,km\,s^{-1}\,pc^2})^{-1}$ \citep{Zavala2022}. 
Adopting a larger $\alpha_{\rm CO}$ would increase $N_\mathrm{H_2}$ and thus decrease $N_{\rm HF}/N_{\rm H_2}$, so our qualitative conclusions below remain unchanged.

\begin{figure}[tbh]
\centering
\includegraphics[width=\columnwidth]{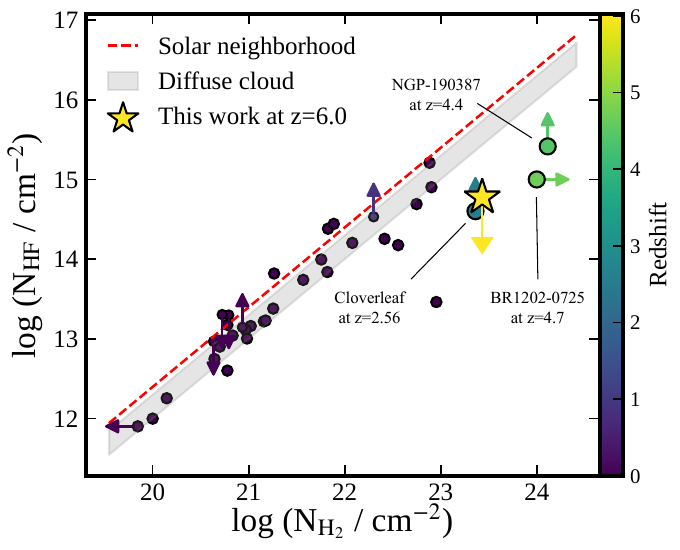}
\caption{HF column density versus H$_2$ column density. Literature measurements of HF in absorption and emission (circles; \citealt{Neufeld2010A&A...518L.108N, Phillips2010A&A...518L.109P, Sonnentrucker2010A&A...521L..12S, Monje2011ApJ...734L..23M, Emprechtinger2012ApJ...756..136E, Kamenetzky2012ApJ...753...70K, Indriolo2013ApJ...764..188I, Pereira-Santaella2013ApJ...768...55P, Kawaguchi2016ApJ...822..115K, Lu2017ApJS..230....1L, Kavak2019A&A...631A.117K}) are color-coded by redshift, and the new measurement from this work is shown as a star.
Arrows indicate upper and lower limits. The red dashed line denotes the fluorine abundance in the solar neighborhood ($\mathrm{F/H}=2.5\times 10^{-8}$; \citealt{Maiorca2014ApJ...788..149M}), and the black dotted line shows the typical $N_{\rm HF}/N_{\rm H_2}$ value found in Galactic diffuse molecular clouds ($(1-2)\times 10^{-8}$; e.g.,~\citealt{Sonnentrucker2010A&A...521L..12S}). 
Three high-redshift sources are highlighted for comparison: the DSFG NGP-190387 at $z=4.4$ \citep{Franco2021NatAs...5.1240F}, the QSO BR1202$-$0725 at $z=4.7$ where HF is detected in emission \citep{Lehnert2020A&A...641A.124L}, and the QSO Cloverleaf at $z=2.56$ \citep{Monje2011ApJ...742L..21M}. 
\label{fig:NHF_NH2}}
\end{figure}

Because HF forms efficiently via the exothermic reaction \(\mathrm{F}+\mathrm{H_2}\rightarrow\mathrm{HF}+\mathrm{H}\) and is resistant to destruction under typical interstellar conditions, it is expected to be the dominant gas–phase reservoir of fluorine wherever hydrogen is predominantly molecular \citep[e.g.,][]{Zhu2002ApJ...577..795Z}. Consequently, the abundance ratio \(N_{\rm HF}/N_{\rm H_2}\) provides a sensitive empirical indicator of fluorine enrichment. 
Figure~\ref{fig:NHF_NH2} places our constraint in this context by comparing it with HF detections and limits in the Milky Way, nearby galaxies, and high–redshift galaxies/quasars. 
In the solar neighborhood, the fluorine abundance is $\mathrm{A(F)} = 12 + \log({\rm F/H})=4.40$, corresponding to $\mathrm{F/H}=2.5\times10^{-8}$ \citep[number ratio,][]{Maiorca2014ApJ...788..149M},\footnote{An earlier determination based on a different excitation potential of the vibrational–rotational HF line reported $\mathrm{A(F)}=4.56$, corresponding to $\mathrm{F/H}=3.6\times10^{-8}$ \citep{Asplund2009ARA&A..47..481A}.}
which provides an approximate upper bound on $N_{\rm HF}/N_{\rm H_2}$ if nearly all fluorine is locked in HF. 
Observations of diffuse molecular clouds in the Milky Way yield slightly lower ratios, $N_{\rm HF}/N_{\rm H_2}\sim (1-2)\times10^{-8}$ \citep[e.g.,][]{Sonnentrucker2010A&A...521L..12S}. 
At low molecular columns ($N_{\rm H_2}\lesssim10^{22.5}\,{\rm cm^{-2}}$), these measurements follow a relatively tight correlation between $N_{\rm HF}$ and $N_{\rm H_2}$.
At higher $N_{\rm H_2}$ and toward higher redshift \citep{Monje2011ApJ...742L..21M, Lehnert2020A&A...641A.124L, Franco2021NatAs...5.1240F}, the relation becomes less clear, showing larger scatter and a possible flattening around $N_{\rm HF}\sim 10^{15}\,{\rm cm^{-2}}$. 
Our $5\sigma$ upper limit for G09.83808 also lies below these reference values, extending fluorine abundance constraints to $z=6$.

\section{Discussion and Conclusion} \label{sec:discussion}
To interpret our HF constraint, we compare it with the galactic chemical evolution model of \citet{Kobayashi2020ApJ...900..179K}, which is also adopted by \citet{Franco2021NatAs...5.1240F}. 
This model self-consistently includes fluorine production from all plausible sources, namely AGB stars (including super-AGB stars), core-collapse supernovae (including hypernovae), and rotationally enhanced WR winds; the neutrino process in \citet{Kobayashi2011ApJ...739L..57K} is uncertain and is {\it not} included.
Therefore, fluorine is enhanced in AGB and WR stars by He-burning. 
The inclusion of the rotating massive stars is a key feature, as stellar rotation enhances internal mixing and mass loss, enabling the ejection of fluorine even at lower stellar masses ($\gtrsim 20M_\odot$) and metallicities.

The yields are taken from \citet{Limongi2018ApJS..237...13L}, adopting a rotation velocity of $v_\mathrm{rot}=300~\mathrm{km~s^{-1}}$ for all stars with initial masses of 13--120~$M_\odot$; 
this represents a maximum WR contribution, since such a high rotation speed is applied uniformly across all massive stars regardless of metallicity.
The fiducial model adopts a Kroupa IMF (a slope of $x=1.3$ for $0.01$--$120~M_\odot$); a more bottom-heavy IMF would increase the relative contribution from AGB stars but suppress that from massive stars, leading to lower total fluorine yields (for a given star formation history (SFH)).
In galactic chemical evolution models, the metallicity in the ISM is calculated as a function of time, assuming the formation epoch, the star-formation timescale $\tau_{\rm SF}$, and the gas-infall timescale $\tau_{\rm in}$. These model parameters are degenerate and should be constrained from observations, which we do in this paper.
For G09.83808, we have two independent observational constraints, namely the gas-phase metallicity ($Z\approx 0.5$–$0.7\,Z_\odot$; \citealt{Tadaki2022}) and the molecular gas fraction ($f_{\rm gas}\simeq 0.20$; \citealt{Zavala2022}). 
Once the observed gas fraction is imposed, the formation epoch is effectively fixed; in the fiducial solution it corresponds to $z_{\rm form}\sim 8.5$.  
Adopting $\tau_{\rm in}=1$ Gyr, the observed metallicity is then reproduced with a star-formation timescale of $\tau_{\rm SF}=0.37$ Gyr.  We refer to this combination as our fiducial model.
These constraints represent a major improvement over the previous study by \citet{Franco2021NatAs...5.1240F}, for which the star-formation history remained largely uncertain. 
We note that in this fiducial model, the newly produced F from AGB stars is only 0.23\% of the total fluorine abundance at $z=6.02$, indicating that the contribution from AGB stars is negligible at this epoch.

\begin{figure*}[tbh]
\centering
\includegraphics[width=0.8\textwidth]{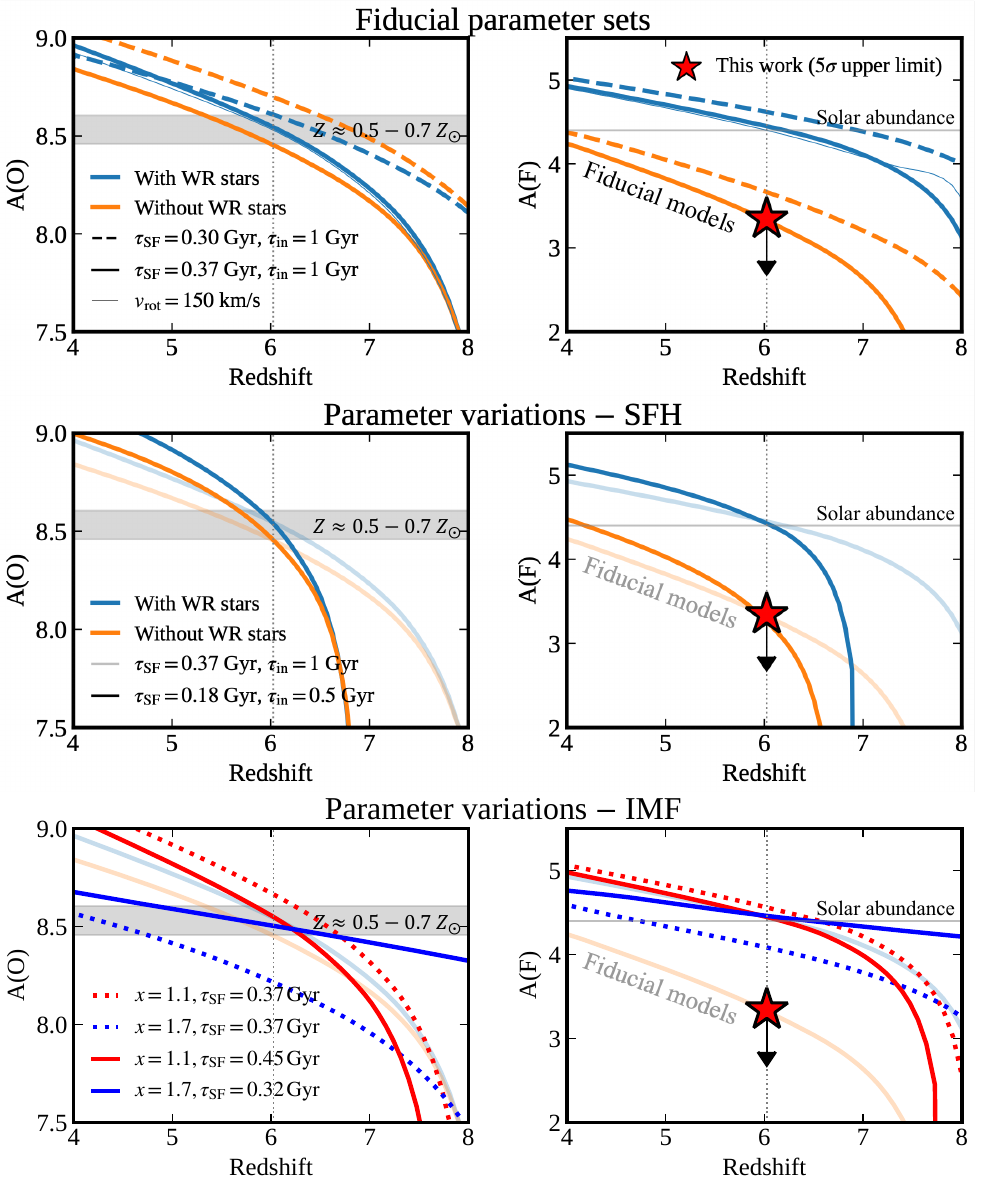}
\caption{Predicted evolution of the oxygen abundance A(O) (left) and fluorine abundance A(F) (right) for chemical–evolution models constrained to reproduce the observed metallicity of G09.83808 ($Z \approx 0.5$–$0.7\,Z_\odot$; \citealt{Tadaki2022}), shown as the gray band in the A(O) panel and its gas fraction ($f_{\rm gas} = 0.2$; \citealt{Zavala2022}).
(Top) Fiducial models, all of which assume a Kroupa IMF with slope $x=1.3$. WR–included cases adopt a stellar rotation velocity of $v_{\rm rot}=300~{\rm km\,s^{-1}}$ except for the thin line.
Blue and orange curves denote models with and without WR yields, respectively.
The solid curves show the model with $\tau_{\rm SF}=0.37$ Gyr and $\tau_{\rm in}=1$ Gyr, while the dashed curves adopt $\tau_{\rm SF}=0.3$ Gyr.
A slower–rotation WR model ($v_{\rm rot}=150~{\rm km\,s^{-1}}$, blue thin curve) is shown for comparison and differs only slightly from the $300~{\rm km\,s^{-1}}$ WR model.
The vertical dotted line marks $z=6.024$.
In the A(F) panel, the gray horizontal line shows the solar abundance \citep{Maiorca2014ApJ...788..149M}, and the red star marks our $5\sigma$ upper limit.
This limit is consistent with models without WR enrichment and with longer star-formation timescales.
(Middle) Models with different star-formation and infall timescales. The bold curves show an alternative SFH ($\tau_{\rm SF}=0.18$ Gyr, $\tau_{\rm in}=0.5$ Gyr) that also satisfies the observed metallicity and gas–fraction constraints, while the faint curves are the fiducial models. To satisfy the constraints, the models yield nearly identical A(F) regardless of the adopted timescales.
(Bottom) Models exploring variations in the IMF slope.
The dotted lines adopt IMF slopes of $x=1.1$ (red) and $x=1.7$ (blue) while keeping all other parameters at the fiducial values; these models fail to reproduce the observed metallicity.
The solid red and blue curves retune the star–formation timescale to $\tau_{\rm SF}=0.45$ and $0.32$ Gyr, respectively, so that the models match the observed metallicity, in which case the predicted A(F) remains similar to the fiducial case.
\label{fig:model}}
\end{figure*}

Figure~\ref{fig:model} (top panel) compares our observational constraint with the fiducial model predictions.
The left panel shows the evolution of oxygen abundance A(O), which serves as a proxy for overall metallicity and is used to anchor each model track to the observed value for G09.83808.  
The right panel shows the corresponding fluorine abundance A(F).  
Our $5\sigma$ upper limit on HF/H$_2$ ($<2.2\times10^{-9}$) is consistent with the models without WR yields and does not require additional fluorine production from massive stars.  

To assess the robustness of this result, we explore variations in the model parameters, as illustrated in the Figure~\ref{fig:model}.
First of all, reducing the rotation velocity of massive stars from 300 to 150\,km\,s$^{-1}$ produces negligible changes in both the oxygen and fluorine evolution tracks (top panel).  
Secondly, We examine models with different star-formation and infall timescales (middle panel).
When the observed metallicity and gas fraction are imposed, changing $\tau_{\rm SF}$ simply requires a compensating change in $\tau_{\rm in}$, and all such combinations converge to the same fluorine abundance.  Thus the predicted A(F) is insensitive to variations in the SFH under these constraints.
Finally, we vary the IMF slope (bottom panel). Changing the Kroupa IMF slope from $x=1.3$ to either $x=1.1$ (top–heavy) or $x=1.7$ (bottom–heavy) yields modest differences in fluorine abundance ($\sim$0.1–0.4\,dex), but the resulting oxygen abundance A(O) fails to match the observed metallicity.
After adjusting $\tau_{\rm SF}$ to reproduce $Z\approx0.5$--$0.7\,Z_\odot$, the predicted A(F) converges back to the fiducial track. 
Taken together, these tests demonstrate that the low HF abundance in G09.83808 cannot be explained with WR stars, even varying stellar rotation velocity, SFH, or IMF slope.  With both the metallicity and gas fraction independently known, the fluorine abundance becomes a direct probe of the underlying nucleosynthesis, allowing us to place strong constraints on the contribution of WR stars at $z>6$.

We discuss possible explanations for the low HF abundance observed in G09.83808.
First, partial depletion of HF onto dust grains could in principle lower the observed HF column density, particularly in dense, cold environments where freeze-out becomes efficient \citep[e.g.,][]{vanderWiel2016A&A...593A..37V}.
However, the relatively high dust temperature ($T_\mathrm{dust}=51\pm4$\,K; \citealt{Tadaki2022}) and the elevated cosmic microwave background temperature at $z=6$ ($T_\mathrm{CMB}=19.2$\,K) render freeze-out inefficient under typical ISM densities ($n_\mathrm{H}\sim10^3$--$10^4$\,$\mathrm{cm^{-3}}$). 
Following the arguments of \citet{Franco2021NatAs...5.1240F}, such high dust and radiation temperatures significantly shorten the thermal desorption timescale relative to the freeze-out timescale, making it unlikely that a significant fraction of HF is locked onto dust grains. 

Second, the galaxy may be observed during a particular phase in an intermittent SFH, a scenario proposed to explain elevated N/O ratios in high-redshift galaxies \citep[e.g.,][]{Kobayashi2024ApJ...962L...6K}.
In this framework, after the initial starburst, secondary gas inflow dilutes the ISM and triggers a secondary star burst, from which WR stars cause high (N, F)/O. Then,  SNe~II dominate to decrease (N, F)/(O, Fe) ratios until the onset of AGB enrichment.
In fact, Figure~4 of \cite{Kobayashi2024ApJ...962L...6K} shows that [F/Fe] declines sharply during this post-burst phase.
If G09.83808 is in the SNe~II enrichment phase, its high metallicity and low fluorine abundance would be naturally explained.

Third, the WR fluorine yields themselves remain highly uncertain, depending on model assumptions such as mass-loss rates, convection, magnetic field, and the way to include rotation \citep[e.g.,][]{Limongi2018ApJS..237...13L}.

By contrast, the $z=4.4$ galaxy studied by \citet{Franco2021NatAs...5.1240F} exhibits a much higher HF abundance.
Within the similar chemical–evolution framework, such a value can be reproduced either by including rotation-enhanced WR yields or by adopting a very short star-formation timescale ($\tau_{\rm SF}\sim0.1\,\mathrm{Gyr}$).
However, due to the lack of independent constraints on the galaxy’s metallicity or SFH, the model remained degenerate in that case. Our result for G09.83808 breaks this degeneracy and reveals that WR-driven fluorine production was not yet efficient for this case. The discrepancy between these two galaxies may reflect diversity in star-formation modes, enrichment histories, or WR activity among early galaxies. Expanding the sample of HF absorption measurements in high-redshift galaxies with known metallicities will be crucial to trace the onset and evolution of fluorine production from massive stars across cosmic time.

\begin{acknowledgments}
We thank the anonymous referee and the editor for their constructive and insightful comments, which helped improve the clarity and quality of this manuscript.
Data analysis was in part carried out on the Multi-wavelength Data Analysis System operated by the Astronomy Data Center (ADC), NAOJ. This research was supported by FoPM, WINGS Program, the University of Tokyo. AT acknowledges the support by JSPS KAKENHI Grant Number JP24KJ0562. CK acknowledge funding from the UK Science and Technology Facilities Council through grant ST/Y001443/1. KK acknowledges the support by JSPS KAKENHI Grant Numbers JP22H04939, JP23K20035, and JP24H00004. 
FM is supported by JSPS KAKENHI grant No. JP23K13142.  
YN acknowledges support from JSPS KAKENHI Grant Numbers JP23K13140 and JP23K20035.
This paper makes use of the following ALMA data: ADS/JAO.ALMA \#2023.1.01281.S. ALMA is a partnership of ESO (representing its member states), NSF (USA) and NINS (Japan), together with NRC (Canada), NSTC and ASIAA (Taiwan), and KASI (Republic of Korea), in cooperation with the Republic of Chile. The Joint ALMA Observatory is operated by ESO, AUI/NRAO and NAOJ. 
\end{acknowledgments}

\begin{contribution}
AT led the project, including the proposal preparation, data analysis, and manuscript writing. 
CK constructed the chemical evolution models.
CK, YY, KK, KT and FM provided overall support throughout the project. 
All authors contributed to discussions and the improvement of the manuscript.

\end{contribution}

%
\facilities{ALMA}

\software{Astropy \citep{astropy:2022},
          Numpy \citep{numpy2020},
          Scipy \citep{2020SciPy-NMeth},
          Matplotlib \citep{matplotlib_Hunter:2007},
          CASA \citep{CASA2022},
          }





\bibliography{reference}{}
\bibliographystyle{aasjournalv7}



\end{document}